\documentstyle[12pt,fleqn]{article}
\begin{document}
\hyphenation{anti-fermion}
\baselineskip = 0.33 in
\begin{center}
\begin{large}
 {\bf {  A NEW INTERPRETATION OF BETHE ANSATZ 
         SOLUTIONS FOR MASSIVE THIRRING MODEL }}

\end{large}

\vspace{1cm}

T. FUJITA\footnote{e-mail: fffujita@phys.cst.nihon-u.ac.jp}
, Y. SEKIGUCHI and K. YAMAMOTO

Department of Physics, Faculty of Science and Technology  
  
Nihon University, Tokyo, Japan

\vspace{1cm}

{\large ABSTRACT} 

\end{center}

We reexamine  Bethe ansatz solutions of the massive Thirring model. 
We solve equations of  periodic boundary conditions numerically 
without referring to the density of states.  It is found  
  that there is only one bound state in the 
massive Thirring model.  The bound state spectrum obtained here 
 is consistent with 
Fujita-Ogura's solutions of the infinite momentum frame prescription.  
 Further, it turns out that there exist no solutions for 
 $string-$like configurations. 
 Instead, we find boson boson scattering states 
 in $ 2-$particle $2-$hole configurations 
 where all the rapidity variables 
 turn out to be real.  

\vspace{1cm}

\vspace{2cm}
Manuscript pages 36: Figures 6: Table 4

\newpage

Proposed running head: " Bethe ansatz for massive Thirring" \par
Mailing Address: \par
T. Fujita \par
Department of Physics \par
Faculty of Science and Technology \par
Nihon University \par
Kanda-Surugadai \par
Tokyo, Japan \par
\vspace{1cm}
Telephone: (03)3259-0887 (Japan) \par
Fax: (03)3259-0887 (Japan) \par
e-mail: fffujita@phys.cst.nihon-u.ac.jp \par
\vspace{3cm}

\begin{enumerate}
\item{\large Introduction}

The sine-Gordon field theory or the massive Thirring model is believed 
to be solved exactly. In their classic paper, Dashen, Hasslacher and 
Neveu presented their solutions to the quantum 
 sine-Gordon model [1].  Although they use 
semiclassical approximations, they consider their solutions to be $exact$. 
This is because their solutions seem to have 
 proper weak and strong coupling limits. 
Also, they showed that 
the sine-Gordon model has a rich spectrum of the charge zero sector. 
This spectrum is translated into the massive Thirring model and 
the bound state mass $ {\cal M} $ (vector boson)  is written as 
$$ {\cal M} = 2m \sin {\pi\over 2} {n\over{(1+{2g_0\over{\pi}}) }} 
 \eqno{(1.1)} $$
where $n$ is an integer and runs from 1 to $(1+{2g_0\over{\pi}}) $. 
 $m$ is the fermion mass of the massive Thirring model. $g_0$ is the 
coupling constant with Schwinger's normalization. 

Further, this spectrum is confirmed by the Bethe ansatz solution [2]. 
This was 
very important since the Bethe ansatz wave function is indeed exact. 
In their paper, Bergknoff and Thacker presented their solutions 
of the massive Thirring model based on the $string$ hypothesis when 
they solve equations of the periodic boundary conditions (PBC) from 
Bethe ansatz wave functions. 

In this way, the massive Thirring model has been considered to be 
solved exactly and is supposed to possess many bound states. 

However, Fujita and Ogura [3] have recently presented their solutions 
of the massive Thirring model employing infinite momentum frame 
prescription. Their spectrum is quite different from eq. (1.1).  
There is only one bound state. However, the bound state energy is 
rather close to the lowest energy of eq.(1.1). The deviation is 
about  10 $\sim$ 20 \% from each other depending on the coupling constant. 
The boson mass  $ {\cal M} $  is given as 
$$ { \tan \alpha  \over{{\pi\over 2}-\alpha}} = 
{g\over{\pi}} \left[ 1+{1\over
{\cos^2 \alpha}}(1-{g\over 4\pi}) \right] \eqno{(1.2)} $$
where   
the boson mass  $\cal M$ is related to $\alpha$  as,
$$ {\cal M} = 2m \cos \alpha  $$ 
where $\alpha$ is between 0 and $\pi\over 2$.  
$g$ is a coupling constant of the massive Thirring model 
with Johnson's normalization [4]. Here, one can easily check 
that there is only one bound state. 

Since this eigenvalue equation is obtained with fermion  
antifermion Fock 
space only, one may say that this is a good approximate solution to the 
massive Thirring model. However, it turns out that the solution eq.(1.2) 
 has all the proper behaviors of the weak and strong coupling limits. 
Instead, if one checks eq.(1.1) carefully, then one sees that 
the semiclassical result of eq.(1.1)  
 does not have a proper weak coupling limit. There, the important point 
 is that  one has to 
take into account  current regularizations in a correct way [3]. 

Further, Ogura, Tomachi and Fujita [5] estimated the effect of 
higher fermion antifermion  
Fock spaces ( two fermion two antifermion Fock space ) 
 and proved that the interactions between two bosons are 
always repulsive. Therefore, it is confirmed that 
 there is only one bound state in the massive 
Thirring model from the infinite momentum frame prescription. 
 
Here, a serious question arises.  How about the Bethe ansatz solution 
for the massive Thirring model ? The Bethe ansatz wave function is 
well known to be exact.  
  This is a strong reason why people 
have believed for almost two decades 
that the bound state spectrum obtained from 
the semiclassical approximation is exact in spite of the fact that they took 
into account only the lowest quantum fluctuations in the path integral. 

In this paper, we reexamine the Bethe ansatz solutions for the massive 
Thirring model and discuss problems  in the treatment by 
Bergknoff and Thacker [2]. 
In particular, we show that the $string$ configurations taken by Bergknoff 
and Thacker  do not satisfy the PBC equations.  
The reason why they have to introduce the $string$ picture is because they 
solve the PBC equations for the density of states. Therefore, they 
could not determine proper rapidities for the positive energy particles. 

It is now clear what one should do. One should solve the PBC equations 
directly for the rapidities ( momenta ) 
without referring to the density of states. 
This is what we have done in this paper. We have solved the PBC equations 
numerically. We consider  a few hundred  particles  to a few thousand 
 particles to make a vacuum. Then, we make one particle-one hole pairs, 
two particle two hole pairs and so on. It is found that there is only 
one bound state for one particle-one hole ($1p-1h$) 
configuration. There is no 
bound state for two particle two hole cases. Further, the bound state 
energy calculated from the Bethe ansatz PBC equations 
turns out to be consistent with that of Fujita-Ogura's solution [eq. (1.2)]  
though we can solve only a limited region of the coupling constant. 

Further, we find the boson boson scattering states in $2p-2h$ configurations.  
Here, it is important to note that the boson boson scattering states have 
rapidity variables which are all real, or at least 
if at all exist, a very small imaginary part.  
Therefore, there is no $string-$like 
solution which satisfies the PBC equations.  

Therefore, we present some evidences that there is only one bound state in 
the massive Thirring model and that the semiclassical result 
by Dashen et  al. is indeed an approximate solution.   

In order that the paper can be understood in a better fashion, we 
 comment that the interactions between particles in the 
massive Thirring model are always repulsive for the positive value 
of the coupling constant. This is trivial but quite important to 
understand the bound state problem of the massive Thirring model. 

The paper is organized as follows. In the next section, we briefly 
explain the Bethe ansatz solution of the massive Thirring model. 
Then, section 3 treats the periodic boundary condition of the 
solution. Also, the regularization in this model 
is discussed.  In section 4,  numerical results of the bound state 
spectrum are presented. In particular, we treat the mass of 
the boson which is the only bound state in this model. 
 In section 5, we 
discuss the boson boson scattering states in $2p-2h$ configurations. 
Section 6 summarises what we have understood and clarified from 
this work. 

\newpage

\item{\large Massive Thirring model and Bethe ansatz solutions}

The massive Thirring model is a 1+1 dimensional field theory with 
current current interactions [6]. Its lagrangian density can be written as 
$$  {\cal L} =  \bar \psi ( i \gamma_{\mu} \partial^{\mu} - m_0 ) \psi 
  -{1\over 2} g_0 j^{\mu} j_{\mu}   \eqno{ (2.1)} $$
where the fermion current $  j_{\mu} $  is written as
$$  j_{\mu} = :\bar \psi  \gamma_{\mu} \psi :   . \eqno{ (2.2)}   $$
Choosing a basis where $\gamma_5$ is diagonal, the hamiltonian is written 
$$  H = \int dx \left[-i(\psi_1^{\dagger}{\partial\over{\partial x}}\psi_1
-\psi_2^{\dagger}{\partial\over{\partial x}}\psi_2 )+
m_0(\psi_1^{\dagger}\psi_2+\psi_2^{\dagger}\psi_1 )+
2g_0 \psi_1^{\dagger}\psi_2^{\dagger}\psi_2\psi_1 \right]  .  \eqno{(2.3)} $$
Now, we define the number operator $N$ as 
$$ N=\int dx \psi^{\dagger}\psi   .  \eqno{(2.4)} $$
This number operator $N$ commutes with $ H$. Therefore, when we construct 
physical states, we must always consider physical quantities with the same 
 particle number $N$ as the vacuum. 
For different particle number state, the vacuum is 
different and thus the model space itself is different. 

The hamiltonian eq.(2.3) can be diagonalized by the Bethe ansatz wave 
functions. Here, we do not repeat the way to construct the Bethe ansatz 
wave functions since it is very well written in Thacker's review  
paper [7].   Therefore, for detail discussions, 
 the reader should refer to his paper. 

The Bethe ansatz wave function $ \Psi (x_1,...,x_N)$ 
for $N$ particles can be written as 
$$ \Psi (x_1,...,x_N)= \exp(im_0 \sum x_i \sinh \beta_i ) 
\prod_{1\leq i < j \leq N} \left[1+i\lambda (\beta_i,\beta_j) \epsilon (x_i-x_j) \right] 
\eqno{(2.5)} $$
where $\beta_i$ is related to the momentum $k_i$ 
 and the energy $E_i$ of $i$-th particle as 
$$ k_i= m_0 \sinh \beta_i   .  \eqno{(2.6a)} $$
$$ E_i= m_0 \cosh \beta_i   .  \eqno{(2.6b)} $$
where $\beta_i$'s are complex variables. 

$ \epsilon (x)$ is a step function and is defined as 
$$ 
\epsilon (x) =  
\left \{
\begin{array}{rc}
-1 & x<0 \\
\ \ \\
1 & x>0 .
\end{array}
\right .
    \eqno{(2.7)} $$ 
$ \lambda (\beta_i,\beta_j)$ is related to the phase shift function 
$ \phi (\beta_i-\beta_j)$ as 
$$  { {1+i \lambda (\beta_i,\beta_j)}\over{1-i \lambda (\beta_i,\beta_j)}}=
 \phi (\beta_i-\beta_j)  .\eqno{(2.8)} $$ 
The phase shift function $\phi (\beta_i-\beta_j)$ can be explicitly written as 
$$ \phi (\beta_i-\beta_j)= -2\tan^{-1} \left[{1\over 2}g_0 \tanh {1\over 2}
(\beta_i-\beta_j) \right]    . \eqno{(2.9)} $$ 
In this case, the eigenvalue equation becomes 
$$ H \mid \beta_1...\beta_N > = (\sum_{i=1}^N m_0 \cosh \beta_i ) \mid 
 \beta_1...\beta_N > \eqno{(2.10)} $$ 
where $\mid \beta_1...\beta_N >$ is related to $\Psi(x_1,...,x_N)$ as  
$$ \mid \beta_1...\beta_N >=\int dx_1...dx_N \Psi(x_1,...,x_N) 
\prod_{i=1}^N  \psi^{\dagger}(x_i,\beta_i) \mid 0 >  . \eqno{(2.11)} $$ 
Also, $\psi (x,\beta)$ can be written in terms of $\psi_1 (x)$ 
and $\psi_2 (x)$ as, 
$$ \psi (x,\beta) = e^{\beta\over 2} \psi_1(x)+ e^{-{\beta\over 2}} 
\psi_2 (x)   .  \eqno{(2.12)}     $$ 
From the definition of the rapidity variable $\beta_i$'s, one sees that 
for positive energy particles, $\beta_i$'s are real while for 
negative energy particles, $\beta_i$ takes the form $i\pi -\alpha_i$ 
where $\alpha_i$'s are real. Therefore, in what follows, we denote 
the positive energy particle rapidity by $\beta_i$ and the negative 
energy particle rapidity by $\alpha_i$. 

\newpage
\item{\large Periodic Boundary Conditions and Regularizations}

The Bethe ansatz wave functions satisfy the eigenvalue equation [eq.(2.10)]. 
However, they still do not have proper boundary conditions. The 
simplest way to define field theoretical models is to put the theory in a 
box of length $L$ and impose periodic boundary conditions (PBC) on the states. 

Therefore, we demand that $\Psi (x_1,..,x_N)$ be periodic in each argument 
$x_i$. This gives the boundary condition 
$$ \Psi (x_i=0) =  \Psi (x_i = L)   . \eqno{(3.1)} $$ 
This leads to the following PBC equations, 
$$ \exp (im_0 L \sinh \beta_i )= \exp (-i \sum_j \phi (\beta_i -\beta_j) ) 
   .   \eqno{(3.2)} $$ 
Taking the logarithm of eq.(3.2), we obtain 
$$ m_0 L \sinh \beta_i = {2\pi n_i} - \sum_j \phi (\beta_i -\beta_j)  
    .  \eqno{(3.3)} $$ 
where $n_i$'s are integer. These are equations which we should now solve. 

Before going to construct physical states, we discuss 
the regularization of the fermion current. This is somehow a complication  
of the massive Thirring model which people often overlook. 
As Klaiber explained in his paper [8], the Thirring 
model has an ambiguity that comes from current regularizations. For any 
field theories with local gauge invariance, 
there is no ambiguity concerning the 
current regularization since one has to make gauge invariant regularizations. 
If one makes gauge invariant regularizations, then one obtains physical 
quantities which do not depend on the choice of the regularization methods. 
The Thirring model has no local gauge invariance and thus may well have some 
ambiguity that arises from the way  one makes regularizations. 

In the treatment of the Bethe ansatz wave functions, we have made a 
regularization when constructing $\Psi (x_1,..,x_N)$. There, we assume 
the following identity for the step function $\epsilon (x)$, 
$$ {d\over{dx}}\epsilon (x) = 2 \delta (x),  
\qquad  \delta (x) \epsilon (x) =0   .  \eqno{(3.4)} $$   
This regularization involves only space coordinate. This indicates that 
the regularization employed here must be Schwinger's regularization [9 ]. 
For Schwinger's regularization, we denote the coupling constant by $g_0$. 
In this case, the value of $g_0$ varies from $-{1\over 2}$ to $\infty$. 
On the other hand, there are other regularizations. In particular, 
Johnson's regularization is most popular [4]. 
There, the current is regularized 
with the point splitting of space and time in a symmetric fashion. This 
has some advantage in that the  current conservation is preserved.  
 The coupling constants in the two different regularizations (Schwinger 
and Johnson ) are related to each other as follows,  
$$ g_0={2g\over{2-{g\over{\pi}}}}   \eqno{(3.5)} $$ 
where $g$ denotes the coupling constant with 
Johnson's regularization. For Johnson's regularization, 
  as Klaiber states, free fermion 
basis states lead to Johnson's regularization. Therefore, if one works 
in perturbation theory, then one uses automatically 
Johnson's regularization [10]. 

In the Introduction, we mentioned that the semiclassical result 
of eq.(1.1) does not have the right weak coupling limit. 
Dashen et al. expanded eq.(1.1) in terms of $g_0$ for the weak 
coupling limit and obtained the result as 
$$ {\cal M}= m \left( 2-g_0^2 +{4\over{\pi}}g_0^3 + .. \right) . 
\eqno{(3.6)} $$
On the other hand, the perturbative calculation of the bound state 
spectrum is found 
$$ {\cal M}= m \left( 2-g^2 +{4\over{\pi}}g^3 + .. \right) . 
\eqno{(3.7)} $$
The important point is that the perturbative treatment automatically 
employs Johnson's regularization and thus the result should be 
written by the coupling constant  $g$ in eq.(3.7). Thererfore, 
it is clear that eq.(3.6) does not agree with eq.(3.7) and thus 
eq.(1.1) does not have the right behavior of the weak coupling 
limit at the order of $O(g^3)$. 
The detailed discussions can be found in ref.[5]. 

Throughout this paper, we use Schwinger's normalization $g_0$.  
Later in this paper, we often specify the table and 
 the figures by the coupling constant 
$g\over{\pi}$ ( Johnson's normalization ). This is because it is easier 
to compare the numerical results with Fujita-Ogura's solutions. But 
the value of $g\over{\pi}$ is easily converted into $g_0$ by eq. (3.5).

\vspace{3cm}
\item{\large Numerical Solutions}

Now, we are ready to construct physical states. The parameters we have here 
are the box length $L$ and the particle number $N$. 
 In this case, the density of the system $\rho$ becomes  
$$ \rho = {N \over{L}}    .  \eqno{(4.1a)}  $$ 
Here, the system is fully characterized by
the density $\rho$. For later convenience, we define the effective 
density $\rho_0$ as   
$$ \rho_0 = {N_0\over{ L_0}}  \eqno{(4.1b)} $$ 
where  $L_0$ and $N_0$ are defined as $L_0 = m_0 L$ and 
 $N_0 = {1\over 2}(N-1)$, respectively.  

\vspace{1cm}
\begin{enumerate}
\item{\large Vacuum state}

First, we want to make a vacuum.  We write the PBC equations 
for the vacuum which is filled with negative energy particles 
( $\beta_i=i\pi -\alpha_i$ ),  
$$  \sinh \alpha_i = {2\pi n_i \over{L_0}}  
 - {2\over{L_0}} \sum_{j \not= i} \tan^{-1}\left[{1\over 2}g_0 
\tanh{1\over 2} (\alpha_i -\alpha_j) \right] ,  \qquad  
(i=1,..,N)   . \eqno{(4.2)} $$ 
 Now, $n_i$ runs as 
$$ n_i = 0, \pm 1, \pm 2, ..., \pm N_0    .  $$ 
Therefore, $n_i$ can be replaced by $i$ and thus eq.(4.2) becomes 
$$  \sinh \alpha_i = {2\pi i \over{L_0}} -
 {2\over{L_0}} \sum_{j \not= i} \tan^{-1}\left[{1\over 2}g_0 
\tanh{1\over 2} (\alpha_i -\alpha_j) \right],  \qquad 
(i= 0, \pm 1,..,\pm N_0 ) . \eqno{(4.3)} $$ 
We fix the values of $L_0$ and $N$, and then can solve eq.(4.3). 
This determines the vacuum. In this case, the vacuum energy $E_v$ can be 
written as  
$$ E_v =- \sum_{i=-N_0}^{N_0} m_0 \cosh \alpha_i  . \eqno{(4.4)} $$ 
To describe physical states, we have to renormalize the energy to some physical point. Therefore, $m_0$ itself  does not play 
any important role.  
 
In fig.1, we show how the vacuum and other particle-hole states are 
 constructed. Fig.1a shows the vacuum state. Depending on the value of 
the coupling constant, the shape changes. This has no ambiguity and 
one can also make the vacuum with solving for the density of states.  

\vspace{1cm}
\item{\large $1p-1h$ configuration}
 
Next, we want to make one particle-one hole $(1p-1h)$ 
 state. That is, we take out one 
negative energy particle ($i_0$-th particle) 
 and put it into a positive energy state. 
In this case, the PBC equations become 
\renewcommand{\theequation}{4.5\alph{equation}}
\setcounter{equation}{0}
\begin{eqnarray}
i\neq i_0 \nonumber \\
 & \sinh\alpha_i & = \frac{2\pi i}{L_0}-\frac{2}{L_0}\tan^{-1}
 \left[ {1\over 2}g_0\coth\frac{1}{2}(\alpha_i+\beta_{i_0})\right] 
 \nonumber \\
 & &- \frac{2}{L_0}\sum_{j\neq i,i_0}\tan^{-1}
\left[{1\over 2}g_0\tanh\frac{1}{2}
 (\alpha_i-\alpha_j)\right] \\
 \nonumber \\
i=i_0 \nonumber \\
 & \sinh\beta_{i_0} & = \frac{2\pi i_0}{L_0}+\frac{2}{L_0}\sum_{j\neq i_0}
 \tan^{-1}\left[{1\over 2}g_0\coth\frac{1}{2}(\beta_{i_0}+\alpha_j)\right]   
\end{eqnarray} 
where $\beta_{i_0}$ can be a complex variable as long as it can satisfy 
eqs.(4.5). 

These  PBC equations determine the energy of the one particle-one 
hole states which we denote by $E_{1p1h}^{(i_0)}$,  
$$ E_{1p1h}^{(i_0)}= m_0 \cosh \beta_{i_0} -
\sum_{\stackrel{\scriptstyle i=-N_0}{i\not= i_0}}^{N_0} 
m_0 \cosh \alpha_i  .  \eqno{(4.6)}   $$  
It is important to notice that the momentum allowed for the 
positive energy state must be determined by the PBC equations. Also, 
the momenta occupied by the negative energy particles  are different from 
the vacuum case. 

The lowest configuration one can consider is the case in which one takes 
out $i=0$ particle and puts it into the positive energy state. 
This is shown in fig.1b. This 
must be the first excited state since it has a  symmetry of 
$\alpha_i = - \alpha_{-i}$. We call this state ``symmetric" since 
it has a left-right symmmetry in fig.1b. 

Next, we consider the following configurations 
in which we take out $i= \pm 1, \pm  2 ,..$ particles 
and put them into the positive energy state. This is shown in fig.1c.  
These are configurations we can build up for one particle-one hole 
state. We should note here that 
we cannot make one particle state or one hole state since there the 
particle number is different from the vacuum. Since the hamiltonian 
of the massive Thirring model commutes with particle number operator, 
we should always stay in the same particle number as the vacuum. 
In this sense, we have 
lost a simplest  renormalization point. It would have been ideal 
if we could make one particle scattering (or continuum ) state 
to which we renormalize the physical mass.  
In this case, we could have had a more predictive power to describe 
the mass of the particle hole state. 

Therefore, we should find out another 
way to renormalize our calculated energy. 
Fortunately, we find that the continuum states of the one particle-one hole 
 appear very clearly. Therefore, we can renormalize the physical mass 
to this point as we will discuss it later.  

\vspace{1cm}
\item{\large $2p-2h$ configurations}

In the same way as above, we can make two particle-two hole  $(2p-2h)$ states. 
Here, we take out $i_1-$th and $i_2-$th particles and put them into positive 
energy states.  
The PBC equations for the two particle-two hole states become  
\renewcommand{\theequation}{4.7\alph{equation}}
\setcounter{equation}{0}
\begin{eqnarray}
i\neq i_1, i_2 \nonumber \\
 & \sinh\alpha_i & =  \frac{2\pi i}{L_0}-\frac{2}{L_0}\tan^{-1}
 \left[ {1\over 2}g_0\coth\frac{1}{2}(\alpha_i+\beta_{i_1})\right] 
 \nonumber \\
& & -\frac{2}{L_0}\tan^{-1}
 \left[ {1\over 2}g_0\coth\frac{1}{2}(\alpha_i+\beta_{i_2})\right] 
 \nonumber \\
 & & - \frac{2}{L_0}\sum_{j\neq i,i_1,i_2}\tan^{-1}
\left[{1\over 2}g_0\tanh\frac{1}{2}
 (\alpha_i-\alpha_j)\right] \\
 \nonumber \\
 i=i_1 \ \ \ \nonumber \\
 & \sinh\beta_{i_1} & = \frac{2\pi i_1}{L_0} +\frac{2}{L_0}\tan^{-1}
\left[{1\over 2}g_0\tanh\frac{1}{2}
 (\beta_{i_1}-\beta_{i_2})\right] \nonumber \\
&&+\frac{2}{L_0}\sum_{j\neq i_1,i_2}
 \tan^{-1}\left[{1\over 2}g_0\coth\frac{1}{2}
 (\beta_{i_1}+\alpha_j)\right] \\ 
\nonumber \\
i=i_2 \ \ \ \nonumber \\
 & \sinh\beta_{i_2} & = \frac{2\pi i_2}{L_0} + \frac{2}{L_0}\tan^{-1}
\left[{1\over 2}g_0\tanh\frac{1}{2}
 (\beta_{i_2}-\beta_{i_1})\right] \nonumber \\
&&+\frac{2}{L_0}\sum_{j\neq i_1,i_2}
 \tan^{-1}\left[{1\over 2}g_0\coth\frac{1}{2}(\beta_{i_2}+\alpha_j)\right] .  
\end{eqnarray} 
In this case, the energy of the $2p-2h$ states  $ E_{2p2h}^{(i_1,i_2)}$ 
becomes   
$$ E_{2p2h}^{(i_1,i_2)}= m_0 \cosh \beta_{i_1}+m_0 \cosh \beta_{i_2}
 -\sum_{\stackrel{\scriptstyle i=-N_0}{i\not= i_1,i_2}}^{N_0} 
m_0 \cosh \alpha_i  .  \eqno{(4.8)}   $$  
Here, we note that the symmetric case ( $i_1 = -i_2$ ) gains the energy 
and therefore is lower than other asymmetric cases of $2p-2h$ states. 

Higher particle-hole states are constructed just in the same way as above. 
But it turns out that already two particle two hole states do not give 
any bound states. Therefore, it is not worthwhile  carrying out 
 numerical calculations of  higher 
particle hole states. 

Now, we discuss numerical results of the particle-hole energy. We solve 
the PBC equations by iterations. Since the equations are nonlinear, 
it is nontrivial to solve them by iterations. Indeed, simple-minded 
iteration procedures do not give good convergent solutions. 

\vspace{1cm}
\item{\large A new iteration method}

Here, we briefly explain how we solve the nonlinear 
coupled equations by computor. 
The type of equation we want to solve can be schematically written as  
$$ {\mbox{\boldmath $f$}}= G( {\mbox{\boldmath $f$}} )   \eqno{(4.9)} $$ 
where ${\mbox{\boldmath $f$}}=(f_1, f_2,..,f_N)$ are the $N$ variables 
that should be determined. $G$ is some function. 
Now, we want to solve it by iterations. The simple-minded iteration 
equation we can make is 
$$ {\mbox{\boldmath $f$}}^{(n+1)}= G( {\mbox{\boldmath $f$}}^{(n)} ) 
  \eqno{(4.10)} $$  
where we start from some initial value of ${\mbox{\boldmath $f$}}^{(0)}$. 
However, this seldom gives a convergent result. Here, instead of eq.(4.10), 
we use the following equation,  
$$ {\mbox{\boldmath $f$}}^{(n+1)}= G \left( s{\mbox{\boldmath $f$}}^{(n)}
+(1-s){\mbox{\boldmath $f$}}^{(n-1)} \right)
   \eqno{(4.11)} $$   
where $s$ is a free parameter that should be chosen so that it gives a 
convergent result. Indeed, if we vary the value of $s$ around 
 $s \sim 0.1$, then we get  
good convergent  results. In particular, for the asymmetric $1p-1h$ 
states, the introduction of $s$ is essential. 

Before going to discussions of our calculated spectrum, we should note 
that we use mostly  $g\over \pi$ instead of $g_0$ in the tables and figures 
of the calculations just for convenience as mentioned before. 
Also, in what follows, we 
treat the cases in which there is some possibility for many bound 
states. This corresponds to rather strong coupling regions. 
Therefore, we only focus on the cases with 
 the coupling constant $\frac{g}{\pi}$ which is larger
than 0.8.  
This is because, in this region, the semiclassical calculation 
predicts many bound states. 
Also, there, twice of the first excited state energy (boson mass) is lower 
than a half of the free fermion antifermion mass and thus there is a chance 
of having some bound states of bosons.              

\vspace{1cm}
\item{\large Energy spectrum}
 
In table 1, we show the calculated energies ( raw data ) of $E_v$, 
$E_{1p1h}^{(0)}$, $E_{1p1h}^{(n)}$ and $E_{2p2h}^{(n_1,n_2)}$ 
for several values of the coupling constants with the 
particle number $N=1601$.  Here, we can put $m_0 =1$ without 
loss of generality. 

For  $E_{1p1h}^{(n)}$ and $E_{2p2h}^{(n_1,n_2)}$, 
we show the lowest six states just to see the structure of the 
spectrum. As can be seen from this table, there is a finite jump between 
 $E_v$, $E_{1p1h}^{(0)}$, $E_{1p1h}^{(1)}$. However, the differences 
 among  $E_{1p1h}^{(n)} \quad (n=1,6)$ 
 are always some small number which is just 
 ${2\pi\over{L_0}}$, corresponding to the smallest momentum in this 
 calculation. The same phenomena occur to the $2p-2h$ cases. These 
 states correspond to the continuum states. This situation is better 
 seen if we plot them in the figure. 

In fig.2, we show the spectrum of the $1p-1h$ and $2p-2h$ states at 
the coupling  constant ${g\over \pi}=1.25$  
 as an example (particle number $N$ is 1601, and $L_0$ is 110). 
 One  clearly sees that there is one bound state and 
then continuum (or scattering) states start. The first bunch of 
continuum states correspond to the $1p-1h$ states. Therefore, 
the lowest part of the continuum state can be identified as the 
free fermion and free antifermion state at rest, which should be 
just ${\cal M} = 2m$ where $m$ denotes a physical fermion mass.  
Namely, we now know that we find a 
renormalization point of the physical mass. The $1p-1h$ 
continuum energy should 
start from the free fermion antifermion mass, that is twice the fermion 
mass. Therefore, the physical fermion mass $m$ can be written as  
$$ m = {1\over 2} \left( E^{(1)}_{1p1h}-E_v \right)  .  $$ 
Then, another bunch  of 
continuum states appear which correspond to the $2p-2h$ states. It is 
amusing to notice that the energy of the $2p-2h$ continuum is just twice 
of the $1p-1h$ state energy. This is quite important since this indicates 
that we solve the PBC equations properly. Also, the physical fermion mass 
which we identify by the continuum state of $1p-1h$ configuration is  
indeed justified. 
From the figure, it is clear that the $2p-2h$ 
configurations 
do not have any bound states but they are two fermion two antifermion free 
states.  
Therefore, we can calculate the bound state mass of $1p-1h$ state 
with  respect to the free fermion antifermion masses. 

Further, there is another bunch of continuum states which correspond 
to the boson boson scattering states though these states are 
not shown in fig.2. These boson boson scattering states appear near 
the twice of the boson mass. As  will be discussed later, the boson 
boson scattering states behave quite differently from the two 
fermion  or four fermion scattering states. 

\vspace{1cm}
\item{\large Bound state mass in $1p-1h$ states}

In fig.3, we show the calculated results of excitation energies 
$\Delta E_{1p1h}^{(0)}$ and  $\Delta E_{1p1h}^{(1)}$ 
for ${g\over{\pi}}=0.8 $ ( $g_0 =4.19$ ), 
 ${g\over{\pi}}=1 $ ( $g_0 =6.28$ ) and ${g\over{\pi}}=1.25 $ ( $g_0 =10.5$ )
 cases as the 
 function of $\rho_0 = {N_0\over{m_0 L} }$. Here, $\Delta E_{1p1h}^{(0)}$ 
 and $\Delta E_{1p1h}^{(1)}$ are defined as the $1p-1h$ state energies 
 with respect to the vacuum,  
 $$ \Delta E_{1p1h}^{(0)} =  E_{1p1h}^{(0)} -E_v   \eqno{(4.12a)}  $$  
  $$ \Delta E_{1p1h}^{(1)} =  E_{1p1h}^{(1)} -E_v  .  \eqno{(4.12b)}  $$  
As can be seen from fig.3, the excitation energies    
$\Delta E_{1p1h}^{(0)}$ and  $\Delta E_{1p1h}^{(1)}$  have almost 
the same slope between them if we plot them 
in log-log scale as the function of $\rho_0$. 
This suggests that we can write them as  
 $$ \Delta E_{1p1h}^{(0)} =  A_0 + B_0 \rho_0^{\alpha}  \eqno{(4.13a)}  $$ 
 $$ \Delta E_{1p1h}^{(1)} =  A_1 + B_1 \rho_0^{\alpha}  \eqno{(4.13b)}  $$   
where $A_i, \  B_i \ \  (i=0,1) $ are constant but depend on the coupling 
constant $g_0$. The important point is that  
$\Delta E_{1p1h}^{(0)}$ and  $\Delta E_{1p1h}^{(1)}$  have the same slope 
$\alpha$. Here, the values of the $\alpha$ for 
${g\over{\pi}}=0.8$, 1.0 and 1.25  turn out to be  
 $$  \alpha =0.42 \qquad {\rm for} \quad 
 {g\over{\pi}}=0.8 \eqno{(4.14a)}   $$  
 $$  \alpha =0.45 \qquad {\rm for} \quad  
 {g\over{\pi}}=1.0 \eqno{(4.14b)}   $$  
 $$  \alpha =0.50 \qquad {\rm for} \quad  
 {g\over{\pi}}=1.25   . \eqno{(4.14c)}   $$   
In fig.4, we show the same excitation energies   
$\Delta E_{1p1h}^{(0)}$ and  $\Delta E_{1p1h}^{(1)}$  as the function of 
$\rho_0^{\alpha}$ for ${g\over{\pi}}=0.8$, 1.0 and 1.25  cases. 
From this figure, one sees that calculated points are almost on the 
straight line. Also, one notices that the calculated results are 
consistent with $A_i =0$.  

We note here that $\alpha$ depends on the coupling constant $g_0$ and 
becomes unity when $g_0 \rightarrow \infty$. But it is always 
smaller than unity. 
In this case, we can take the field theory limit $\rho \rightarrow 
\infty $. This can be seen in the following way. Let us take the 
$ \Delta E_{1p1h}^{(0)} $ case, for example. If we write the bare mass  
$m_0$ explicitly, then $\Delta E_{1p1h}^{(0)} $ can be written as   
 $$ \Delta E_{1p1h}^{(0)} = m_0 \left( A_0 + B_0 
 \left( {\rho\over m_0} \right)^{\alpha} \right) .  \eqno{(4.15)}  $$  
Now, we want to let $\rho \rightarrow \infty$, keeping 
$ \Delta E_{1p1h}^{(0)}$ finite. Since $\alpha$ is smaller than unity, 
we can make a fine-tuning of $m_0$ such that  
$$  m_0^{1-\alpha} \rho^{\alpha} = {\rm finite}  . $$  
This means that we should let $m_0 \rightarrow 0 $, and thus the second 
term of eqs. (4.13) becomes dominant in the field theory limit. 
In this case, we can identify the mass of the bound state $\cal M$ as  
$$ {\cal M} = 2m   \lim_{\rho \rightarrow \infty} 
\left( { \Delta E_{1p1h}^{(0)}\over{\Delta E_{1p1h}^{(1)} }} \right) = 
2m {B_0\over{B_1}}   .  \eqno{(4.16)}   $$  
For ${g\over{\pi}}=0.8$, 1.0 and 1.25   
 cases, we find  
$$ B_0= 2.55 , \qquad B_1 = 5.06 \qquad {\rm for} \quad {g\over{\pi}}=0.8
   \eqno{(4.17a)} $$  
$$ B_0= 2.25 , \qquad B_1 = 6.0 \qquad {\rm for} \quad {g\over{\pi}}=1.0
   \eqno{(4.17b)} $$  
$$ B_0= 1.63 , \qquad B_1 = 7.0 \qquad {\rm for} \quad {g\over{\pi}}=1.25 .
   \eqno{(4.17c)} $$  
Thus, the mass of the bound states for ${g\over{\pi}}=0.8$, 1.0 and 1.25 
 becomes   
$$  {\cal M}= 1.01   \qquad {\rm for} \quad 
{g\over{\pi}}=0.8  \eqno{(4.18a)}  $$  
$$  {\cal M}= 0.75   \qquad {\rm for} \quad 
{g\over{\pi}}=1.0  \eqno{(4.18b)}  $$  
$$  {\cal M}= 0.47   \qquad {\rm for} \quad 
{g\over{\pi}}=1.25  \eqno{(4.18c)}  $$  
Here, we comment on the minimum theoretical errors 
which may arise from the minimum 
momentum of the calculation ${2\pi\over{L_0}}$. This gives rise to an error 
for the mass  
$$\Delta {\cal M} \approx {2\pi\over{L_0}}
{2\over{\Delta E_{1p1h}^{(1)} }}2m \approx 0.08 m   . $$  
However, the above errors may not be the largest one. The larger errors 
may well come from the fact that we have still not yet reached sufficiently 
large values of $N$ and  $\rho_0$ such that 
 $\Delta E_{1p1h}^{(0)}$ and  $\Delta E_{1p1h}^{(1)}$  have the same slope 
$\alpha$. 

The calculated boson mass (eqs.(4.18)) 
 should be compared to those predicted by other methods. 
 In table 2, we show the predictions by the infinite momentum frame 
 calculation ( Fujita and Ogura ), 
 the semiclassical method ( Dashen et al.) 
 and the Bethe ansatz solution with $string$ hypothesis ( Bergknoff and 
 Thacker ).  The comparison can be better seen if we plot them in the figure. 
  In fig.5, we show the boson mass predicted by different methods 
  for the whole range of the coupling constant. The black circles are 
  the present calculation.  
  The fig.5 indicates that the present calculation is 
 consistent with that predicted by Fujita and Ogura (the solid line). 
 Also, one can say that the difference 
 between the present result and the semiclassical one (the dashed line) 
  is not very large and is probably within an error of our calculation. 
However, the predicted value of Bergknoff and Thacker (the dashed-dotted line) 
 is very different 
from the present result. This is quite important since we solve the same 
PBC equations as Bergknoff and Thacker, keeping the same coupling 
constant $g_0$. The only difference between the present calculation and 
Bergknoff and Thacker must lie in the treatment of determining the 
rapidity of the positive energy particles. There, they employed the 
$string$ hypothesis since their method that uses the density of states 
cannot determine the rapidity of positive energy particle in the 
particle-hole excitations. This indicates that the $string$ hypothesis 
may not be a very good approximate scheme for the massive Thirring model 
even though it is a good working tool for the nonlinear Schr\"odinger 
model. In ref.[15], it is proved that, in the strong coupling limit, 
the $string$ picture does not satisfy the PBC equation. 
  
Note that Bergknoff and Thacker claim that they reproduced the semiclassical 
results by Dashen et al.. There, when they compare their spectrum with the 
semiclassical result, they have made  a   
renormalization of the coupling constant $g$. However, this is not justified 
since  Dashen et al. use  Schwinger's normalization of the coupling 
constant, namely, the same $g_0$ as Bergknoff and Thacker. 
Here, it is clear since we used Bergknoff-Thacker's 
formula, keeping the same coupling constant $g_0$ as appeared 
in this paper since we solved the same equation with the same 
boundary conditions.  

At this point, we comment on the renormalization of the coupling 
constant. Here, 
we do not have to make any renormalizations of the coupling constant when  
calculating physical quantities.  Only the mass renormalization is 
needed.  The only important point is that one has to know 
which regularization of the current one has employed in his calculation. 

Now, we discuss the limit of our calculation. Unfortunately, 
we cannot find solutions beyond some value of $\rho_0$. 
That is entirely due to the problem of our computer program. 
Until now, we do not find any better way of calculating the cases 
with higher values of $\rho_0$. But as far as the 
${g\over{\pi}}=0.8$, 1.0 and 1.25  cases  
 are concerned, we have reached relatively large value of $\rho_0$ 
even though it is not yet sufficient. This can be seen in the 
following way. In order that the vacuum can be constructed in a proper 
fashion, the following conditions must be satisfied,  
$$  {2\pi\over L} \ll m_0 \ll {2\pi N_0\over L}   .  \eqno{(4.19)}  $$  
For the cases of ${g\over{\pi}}=0.8$, 1.0 and 1.25, we have  
\begin{eqnarray*}
  L_0 &=& m_0 L \approx 10 \sim 100 
\\
 N_0 &=& 800  . 
\end{eqnarray*}  
Therefore, the above equation becomes  
$$  (0.6 \sim 0.06 ) \ll 1 \ll (50 \sim 500) .  \eqno{(4.20)}    $$    
Here, the left-hand inequality is not very well satisfied. 
This corresponds to the large value of $\rho_0$. This may well generate 
larger errors than $ \Delta {\cal M}$. To overcome this difficulty, 
we have to increase the number of particle $N$ by order of magnitude.  
From these considerations, we can say 
that our calculated value of the bound state mass is reliable 
only to some extent to which the slopes of 
$\Delta E_{1p1h}^{(0)}$ and  $\Delta E_{1p1h}^{(1)}$ are the same. 
As mentioned above, the two slopes are still slightly different. 
In particular, the two slopes for ${g\over{\pi}}=1.25$ case are 
still appreciably different from each other. 

In fig.6, we show the calculated results for
  ${g\over{\pi}}= 1.5, \  1.7$. As can be seen from this  
  figure, we have not yet reached sufficiently large values of $\rho_0$.  
Therefore, the slopes of 
the $\Delta E_{1p1h}^{(0)}$ and the $\Delta E_{1p1h}^{(1)}$  are still 
quite different from each other.   Thus, we cannot make 
the field theory limit for these cases.   This is entirely due to 
the poorness of our computer program. Indeed, 
for the coupling constants ${g\over{\pi}}$ larger than 1.8, 
the situation is even worse.  
We can not find any reasonable solutions yet. In this respect, the 
calculations presented here are only very limited. This is certainly 
connected to the fact that the PBC equations are highly nonlinear. 

\end{enumerate}

\vspace{3cm}

\item{\large Excitation energies in $2p-2h$ states} 

As shown in fig.2, the excitation energies of the $2p-2h$ states 
always appear at the energy of four times physical fermion mass. 
This is one of the strong evidences that we solve the PBC equations 
properly. However, the above calculations do not include 
 the important states 
of $2p-2h$ configurations, that is,  solutions 
for boson-boson scattering states near the energy of the 
twice of the boson mass. 

In this section, we present the boson boson scattering states 
which are found in the $2p-2h$ configurations. 

First, we note that the iteration procedure described 
in section 4(d) with real rapiditiy variables always give 
the four fermion states as we saw before. 

Here, we make the rapidity variables all complex so that 
we can find the $string-$like solutions which should correspond 
to the boson boson scattering states. Now, it turns out that, in 
order to find the boson boson states, we must vary the initial 
values of the rapidities in the iteration procedure. Depending 
on the initial values, we find the boson boson scattering states 
in the one case, but find the four fermion states in the other case. 

In any case, once we know how to find the boson boson states, 
then, we can obtain the boson boson state energy for any states 
we need. Here, we note that all the resulting rapidity variables 
are found to be real, even though we start from the complex 
initial values of the rapidity. 

Therefore, this shows that there is no $string-$like solution 
which satisfies the PBC equation. The imaginary part of the 
rapidity, if at all exist, should be quite small. 

In Table 3, we show the first six excitation energies 
 of the boson boson states $E_{2p2h}^{BB}$ 
 for several values of the coupling constants. 
 This clearly shows that these states are scattering states 
 (unbound states) and make a continuum spectrum. 
It is quite interesting to observe that these continuum states 
differ from each other by the energy unit which is smaller than 
the case with four fermion states. In fact, the energy unit of the 
continuum state for the four fermion states is ${4\pi\over L_0}$ 
( see Table 1 ) 
while the energy unit for the boson boson states is a factor of 5  to 10 
smaller than ${4\pi\over L_0}$. 
This suggests that these states 
are indeed quite different from four fermion states. 

Now, it should be fair to note that it is indeed quite difficult 
to find the boson boson states by computor. This is mainly 
because we do not know $a$ $priori$ which of the states $(n,-n)$ 
is the lowest. The lowest energy of the boson boson scattering 
state can be obtained only after we obtain all the boson boson 
state energies. Therefore, it is always very much cumbersome 
to find the lowest energy of the boson boson scattering states. 
This must be connected to the fact that the boson boson scattering 
states are constructed by the two step processes, first by 
making one boson state and then by making the boson boson 
scattering states.  

In Table 4, we summarize the calculated results of the excitation 
energies for several values of the coupling constants as well 
as for the several cases of the particle numbers $N$ 
and the box sizes $L_0$. Here, we plot  the lowest state of 
the boson boson excitation 
energy $\Delta E_{2p2h}^{BB}$  with respect to the vacuum, 
 and the first 
excited state (boson state) $\Delta E_{1p1h}^{(0)}$ 
 as well so that we can compare 
the calculated results of the boson boson state energy    
with twice of the boson mass. Also, for comparison, we show 
 the lowest state of 
the two  fermion scattering  state $\Delta E_{1p1h}^{(1)}$,  
and the lowest state of the four fermion scattering  state 
$\Delta E_{2p2h}^{(1,-1)}$.  As can be seen from the Table 4, 
the boson boson states appear lower than the twice of the boson mass.   
In particular, depending on the density $\rho_0$, the boson boson 
state energy becomes much lower than the twice of the boson mass. 
Also, the dependence of the boson boson state energy on the $\rho_0$ 
is in general quite different from the $1p-1h$ state energies. 
We do not fully understand the reason why there is such a significant 
difference between the boson boson state energy and the free fermion 
 ($1p-1h$ and $2p-2h$ states) energies. 
 
From the present calculations, we get to know that the boson boson 
state energy depends strongly on the box size $L_0$. Unfortunately, 
in the present calculation, the box size is much too small. 
In particular, for large $\rho_0$, the value of $L_0$ is too small 
to satisfy the criteria of eq.(4.19). This may be one of the 
reasons why we obtain the very low excitation energy of the boson 
boson states. 

Since the dependence of 
the boson boson excitation energy $\Delta E_{2p2h}^{(BB)}$ on the density 
$\rho_0$ is quite different from those of $\Delta E_{1p1h}^{(0)}$ and 
$\Delta E_{1p1h}^{(1)}$,  we cannot, therefore,
 make the field theory limit 
as we have made it for $\Delta E_{1p1h}^{(0)}$ and 
$\Delta E_{1p1h}^{(1)}$ cases. 
As stated above,  the values of $N$ and $L$ which we took here 
are still too small in order to obtain some quantitative numbers 
for $\Delta E_{2p2h}^{(BB)}$. In this respect, we only get some 
qualitative pictures of the boson boson scattering states. 

Finally, we want to make a comment on the validity of the 
boson boson state energy. In ref.[15], the strong coupling 
expansion is presented. There, the excitation energy is 
evaluated analytically. This analytic expression is compared 
to our calculations of the boson boson state energy. 
It turns out that the both calculations agree with each other 
quite well for the large values of $L_0$.

\newpage

\item{\large Conclusions and discussions}

We have presented a new interpretation of the Bethe ansatz solutions of 
the massive Thirring model. Here, we solve the PBC equations 
directly but numerically without referring to the density of states 
or $string$ hypothesis. It is found that the Bethe ansatz solutions 
produce one bound state (a boson). This spectrum as the function of 
the coupling constant is consistent with Fujita-Ogura's solution. 

 Also, it is shown that the $string$ 
configurations taken by Bergknoff and Thacker do not satisfy the 
PBC equations and thus their $string$ is not a solution of the PBC 
equations.  In this way, the present result rules out a belief 
that the semiclassical 
result for the massive Thirring model is exact. 

Here, we want to give an intuitive argument why there is only one bound 
state from the PBC equations. The vacuum is represented by fig.1a. 
This figure shows that the left-right symmetry $(\alpha_i= - \alpha_{-i})$ 
is preserved there. Now, we make one particle one hole state 
as shown in fig.1b 
and 1c. The fig.1b has the left-right symmetry and therefore it gains the 
energy. The important point is that there is only one state that has this 
symmetry. In fact, the fig.1c does not have the left-right symmetry and 
therefore its energy is rather high. This obviously gives a continuum 
spectrum. Now, let us consider the two 
particle-two hole states. In fig.1d, we show only the symmetric case since 
the asymmetric configurations do not gain the energy. Here, the important 
point is that there is no special configuration which differs from others. 
The lowest configuration is different from the next 
lowest only by ${4\pi\over L_0}$ and so on. 
Therefore, it is clear that these $2p-2h$ states should describe some 
continuum states. For higher particle-hole states, the situation is the same 
as the $2p-2h$ states. Therefore, it is qualitatively clear why there is only 
one bound state from the PBC equations. This is indeed confirmed by numerical 
calculations. 

Also, the strong coupling expansion is performed in ref.[15] and 
the analytic expressions are obtained for the vaccum state energy 
as well as the boson boson scattering states. 
There, it turns out that the boson boson scattering states 
which are made of continuum states coincide with the twice of 
the boson mass. Therefore, we also learn 
from the strong coupling expansion that the $2p-2h$ states 
do not give any bound states, to say the least, 
 there is no bound state 
found in the analytic expressions of the strong coupling expansion.     

This should naturally bring up many problems and questions 
concerning those methods 
or solutions which agree with the semiclassical results [11,12,13,14,17]. 

First, we comment on Baxter's solution to the Heisenberg XYZ model [11].
This model is shown to be equivalent to the massive Thirring model [12].
Baxter's solution is indeed exact. However, he presented 
his solution  only  
for the largest eigenvalue of the XYZ hamiltonian. After Baxter's solution, 
Johnson, Krinsky and McCoy [13] obtained the next largest eigenvalue by 
emloying the $string$ hypothesis. Again, the same approximate scheme as the 
$string$ is used there. As we showed in this paper, these states constructed 
from the $string$ hypothesis correspond to boson-boson scattering states. 
In quantum mechanical terminology, they may correspond to real quantum states, 
but they are not bound states! Therefore, the solution of the Heisenberg XYZ 
model has the same problem as Bergknoff-Thacker's solution.

Now, we want to discuss the S-matrix method by Zamolodchikov and Zamolodchikov 
[14]. This factorized S-matrix method is also known to give the same spectrum 
as the semiclassical result for the sine-Gordon field theory or the massive 
Thirring model. 
 Concerning the factorization of the S-matrix 
 for the particle-particle scattering 
in the massive Thirring model, 
one may  convince oneself that the
factorization is indeed satisfied. 

However, there is a serious problem for the S-matrix factorization 
of the particle hole scattering. The problem is that the rapidity 
variables determined for $n-$particle $n-$hole states are different 
from each other as well as 
 those determined for the vacuum. Since the Lagrangian of the 
massive Thirring model satisfies the charge conjugation, 
one tends to believe that the crossing symmetry should be automatically 
satisfied. Indeed, the crossing symmetry itself should hold. 
However, we should be careful whether the crossing symmetry can 
commute with the factorization of the S-matrix or not. 
Recent calculations in ref.[16] show that the crossing symmetry 
and the factorization of the S-matrix do not commute with eath other. 
Therefore, it turns out that the S-matrix factorization 
for the particle hole scattering does not hold. In a sense, it is 
reasonable that the S-matrix factorization is consistent with the 
semiclassical results since it is indeed due to the consequence 
of the neglect of the operartor commutability. 

Also, we discuss the results which are obtained from the inverse 
scattering methods [17]. This also gives 
the same spectrum as the semiclassical 
result. In this case, one can start from the soliton and antisoliton 
scattering. Therefore, there is no problem concerning the crossing 
symmetry. However, in this case, one does not know how to quantize 
the solitons and antisolitons in a proper fashion. Therefore, 
it is natural that the inverse scattering treatment remains 
semiclassical for the massive Thirring model 
 and thus the result of the inverse scattering result 
agrees with the spectrum of the WKB method. 

At this point, it should be fair to comment on the physical implication 
of the present result. Though the present paper shows that 
the excitation energy spectrum of sine-Gordon/massive Thirring model 
obtained by the WKB method is not exact, the physical significance 
of this proof is not very great. We show that the boson boson 
states are all scattering states (unbound) while the WKB method 
insists that the lowest one of the boson boson states is a bound state. 
This energy difference is very tiny, and further any physical effects 
of this difference may be of a minor importance. But we simply stress 
that the WKB result is not exact but gives an approximate solution 
to the model. 

There are, however, still many things to be done. In particular, it is 
 very interesting to find analytic solutions of the PBC equations. 
 Up to now, we 
do not know them. The vacuum itself is solved analytically 
with the help of the density 
of states and the vacuum energy is obtained analytically. 
However, it seems still quite difficult to solve analytically 
the PBC equations for particle-hole configurations, though 
we believe that the analytic solution must exist and should be 
found before long. 

Finally, we make a comment as to whether there exist any exact solutions 
of the bound state spectrum in the massive Thirring model. 
Since the semiclassical result turns out to be not exact, it should be 
interesting to check whether Fujita-Ogura solution is exact or not. 
Although this spectrum of eq.(1.2) has all the nice featrures  
of the weak as well as strong 
coupling regions, there is no proof that the solution is exact. 
Therefore, it should be challenging to prove or disprove the exactness 
of the spectrum of eq.(1.2) since this is the only candidate that is 
still left undecided. 

\vspace{1cm}

Acknowledgments: We  thank C. Itoi, F. Lenz, H. Mukaida, A. Ogura  
and R. Woloshyn for stimulating discussions and comments. 
 
\newpage

{\large REFERENCES }
\baselineskip = 8 mm

1.  R. F. Dashen, B. Hasslacher and A. Neveu, 
Phys. Rev. {\bf D11} (1975), 3432 

2. H. Bergknoff and H.B. Thacker, Phys. Rev. Lett. {\bf 42} (1979), 135  

3. T. Fujita and A. Ogura,  Prog. Theor. Phys. {\bf 89} (1993), 23 

4. K. Johnson, Nuovo Cimento, {\bf 20} (1961), 773 

5. A. Ogura, T. Tomachi and T. Fujita, Ann. Phys. (N.Y.) 
{\bf 237} (1995), 12

6. W. Thirring, Ann. Phys. (N.Y) {\bf 3} (1958), 91 

7. H.B. Thacker, Rev. Mod. Phys. {\bf 53} (1981), 253 

8. B. Klaiber, in Lectures in Theoretical Physics, 1967, edited by \par
\qquad A. Barut and W. Britten (Gordon and Breach, NY, 1968) 

9. J. Schwinger, Phys. Rev. Lett. {\bf 3} (1959), 296   

10. A. Ogura, Ph. D thesis, Nihon University (1994)

11. R.J. Baxter, Ann. Phys. (N.Y.) {\bf 70} (1972), 193 

12. A. Luther, Phys. Rev. {\bf B14} (1976), 2153 

13. J.D. Johnson, S. Krinsky and B.M. McCoy, Phys. Rev. {\bf A8} (1973), 2526 

14. A.B.Zamolodchikov and A.B.Zamolodchikov, Ann. Phys. {\bf 120} (1979), 253 

15. T. Fujita, C. Itoi and H. Mukaida, to be published. 

16. T. Fujita and M. Hiramoto, to be published.

17. V.E. Korepin and L.D. Faddeev, Theor. Math. Phys. {\bf 25} (1975), 1039 

\end{enumerate}
\newpage
\begin{center}
\underline{Table 1} \\
\ \ \\
 $N$=1601 \ \ \  $L_0$=100 \ \ \ \ \ \  \\
\ \ \\ 
\begin{tabular}{|l|c|c|c|c|}
\hline
 & \multicolumn{1}{|l}{$\begin{array}{lcl}
   \frac{g}{\pi}& = &1 \\
   \end{array}$} 
 & \multicolumn{1}{|l}{$\begin{array}{lcl}
   \frac{g}{\pi} & = & 1.25 \\
   \end{array}$} 
 & \multicolumn{1}{|l}{$\begin{array}{lcl}
   \frac{g}{\pi} & = & 1.5 \\
   \end{array}$} 
 & \multicolumn{1}{|l|}{$\begin{array}{lcl}
   \frac{g}{\pi} & = & 1.7 \\
   \end{array}$} \\   
\hline
&&&& \\
$E_v$ & $- 9095.31$ & $-6215.70$ & $-4205.83$ & $-2995.13$ \\
\hline
&&&& \\
$E^{(0)}_{1p1h}$ & $- 9089.43$ & $-6210.69$ & $-4201.76$ & $-2991.83$ \\
\hline
&&&& \\
$E^{(1)}_{1p1h}$ & $- 9080.78$ & $-6197.08$ & $-4182.54$ & $-2966.95$ \\
&&&& \\
$E^{(2)}_{1p1h}$ & $- 9080.72$ & $-6197.02$ & $-4182.48$ & $-2966.89$ \\
&&&& \\
$E^{(3)}_{1p1h}$ & $- 9080.66$ & $-6196.96$ & $-4182.42$ & $-2966.82$ \\
&&&& \\
$E^{(4)}_{1p1h}$ & $- 9080.59$ & $-6196.90$ & $-4182.35$ & $-2966.76$ \\
&&&& \\
$E^{(5)}_{1p1h}$ & $- 9080.53$ & $-6196.83$ & $-4182.29$ & $-2966.70$ \\
&&&& \\
$E^{(6)}_{1p1h}$ & $- 9080.47$ & $-6196.77$ & $-4182.23$ & $-2966.64$ \\
\hline
\hline
&&&& \\
$E^{(1,-1)}_{2p2h}$ & $- 9066.23$   & $-6178.55$ & $-4159.13$ & $-2938.79$  \\
&&&& \\
$E^{(2,-2)}_{2p2h}$ & $- 9066.10$   & $-6178.43$ & $-4158.99$ & $-2938.66$  \\
&&&& \\
$E^{(3,-3)}_{2p2h}$ & $- 9065.97$   & $-6178.30$ & $-4158.86$ & $-2938.52$  \\
&&&& \\
$E^{(4,-4)}_{2p2h}$ & $- 9065.85$   & $-6178.17$ & $-4158.74$ & $-2938.38$  \\
&&&& \\
$E^{(5,-5)}_{2p2h}$ & $- 9065.72$   & $-6178.04$ & $-4158.61$ & $-2938.25$  \\
&&&& \\
$E^{(6,-6)}_{2p2h}$ & $- 9065.59$   & $-6177.91$ & $-4158.49$ & $-2938.12$  \\
\hline
\end{tabular}
\\
\vspace*{1cm}  

\begin{minipage}{13cm}
We plot the calculated energies of $E_v$, $E_{1p1h}^{(n)} \  (n=0,6)$ 
and $E_{2p2h}^{(n,-n)} \  (n=1,6)$ for some values of the coupling 
constant $g\over{\pi}$ 
with the fixed $L_0=100$. The number of particles here 
is $N=1601$. Note that we put $m_0 =1$ in our calculations.  
\end{minipage}
\end{center}

\newpage
\begin{center}
\underline{Table 2}  \hspace{1.5cm}  \\
\ \\
\begin{tabular}{|c|c|c|c|c|c|}
\hline
 & & \multicolumn{1}{l|}{Present} & \multicolumn{1}{l|}{Fujita}  
 & \multicolumn{1}{l|}{Dashen et al.} & \multicolumn{1}{l|}{Bergknoff} \\
 & & \multicolumn{1}{l|}{Calculation} & \multicolumn{1}{l|}{Ogura} & 
 & \multicolumn{1}{l|}{Thacker} \\
\hline
&&&&& \\
${\cal M}$ & $\displaystyle{\frac{g}{\pi}=0.8}$ 
& $1.01  m$ & $0.98m$ & $0.83m$ & $0.51m$ \\
& ($g_0=4.19$) &&&& \\
\hline
&&&&& \\
${\cal M}$ & $\displaystyle{\frac{g}{\pi}=1.0}$ 
& $0.75  m$ & $0.77m$ & $0.62m$ & $0.34m$ \\
& ($g_0=6.28$) &&&& \\
\hline 
&&&&& \\
${\cal M}$ & $\displaystyle{\frac{g}{\pi}=1.25}$ 
& $0.47  m$ & $0.54m$ & $0.41m$ & $0.20m$ \\
& ($g_0=10.5$) &&&& \\
\hline 
\end{tabular}

\vspace{2cm}
\begin{minipage}{13cm}
We plot the predicted values of the boson mass $\cal M$ by the present 
calculation, by the infinite momentum frame calculation ( Fujita-Ogura ), 

by the semiclassical method ( Dashen et al.) and by the Bethe ansatz 
technique  with $string$ hypothesis ( Bergknoff - Thacker ). 
\end{minipage}
\end{center}

\newpage
\begin{center}
\underline{Table 3} \\
\ \ \\

$N=1601$, \ \  $L_0=100$ \\

\begin{tabular}{|l||r@{\extracolsep{0pt}.}l|r@{\extracolsep{0pt}.}l|
r@{\extracolsep{0pt}.}l|r@{\extracolsep{0pt}.}l|}
\hline
 & \multicolumn{2}{c|}{ } & 
 \multicolumn{2}{c|}{ } & 
 \multicolumn{2}{c|}{ } & 
 \multicolumn{2}{c|}{ } \\ 
 & \multicolumn{2}{c|}{$\displaystyle{\frac{g}{\pi}=1}$} & 
 \multicolumn{2}{c|}{$\displaystyle{\frac{g}{\pi}=1.25}$} &
 \multicolumn{2}{c|}{$\displaystyle{\frac{g}{\pi}=1.5}$} &
 \multicolumn{2}{c|}{$\displaystyle{\frac{g}{\pi}=1.7}$} \\
 & \multicolumn{2}{c|}{ } & 
 \multicolumn{2}{c|}{ } & 
 \multicolumn{2}{c|}{ } & 
 \multicolumn{2}{c|}{ } \\ 
\hline
 & \multicolumn{2}{c|}{ } & 
 \multicolumn{2}{c|}{ } & 
 \multicolumn{2}{c|}{ } & 
 \multicolumn{2}{c|}{ } \\ 
$E_{2p2h}^{(BB)*1}$ & 
 $-$9089 & 54 &
 $-$6209 & 95 &
 $-$4200 & 24 &
 $-$2989 & 82 \\
 & \multicolumn{2}{c|}{ } & 
 \multicolumn{2}{c|}{ } & 
 \multicolumn{2}{c|}{ } & 
 \multicolumn{2}{c|}{ } \\ 
$E_{2p2h}^{(BB)*2}$ &  
 $-$9089 & 53 &
 $-$6209 & 93 &
 $-$4200 & 22 &
 $-$2989 & 80 \\
 & \multicolumn{2}{c|}{ } & 
 \multicolumn{2}{c|}{ } & 
 \multicolumn{2}{c|}{ } & 
 \multicolumn{2}{c|}{ } \\ 
$E_{2p2h}^{(BB)*3}$ & 
 $-$9089 & 51 &
 $-$6209 & 92 &
 $-$4200 & 21 &
 $-$2989 & 78 \\
 & \multicolumn{2}{c|}{ } & 
 \multicolumn{2}{c|}{ } & 
 \multicolumn{2}{c|}{ } & 
 \multicolumn{2}{c|}{ } \\ 
$E_{2p2h}^{(BB)*4}$ & 
 $-$9089 & 50 & 
 $-$6209 & 90 & 
 $-$4200 & 19 & 
 $-$2989 & 75 \\
 & \multicolumn{2}{c|}{ } & 
 \multicolumn{2}{c|}{ } & 
 \multicolumn{2}{c|}{ } & 
 \multicolumn{2}{c|}{ } \\ 
$E_{2p2h}^{(BB)*5}$ & 
 $-$9089 & 48 &
 $-$6209 & 89 & 
 $-$4200 & 17 & 
 $-$2989 & 73 \\
 & \multicolumn{2}{c|}{ } & 
 \multicolumn{2}{c|}{ } & 
 \multicolumn{2}{c|}{ } & 
 \multicolumn{2}{c|}{ } \\ 
$E_{2p2h}^{(BB)*6}$ & 
 $-$9089 & 47 & 
 $-$6209 & 87 & 
 $-$4200 & 15 & 
 $-$2989 & 71  \\
 & \multicolumn{2}{c|}{ } & 
 \multicolumn{2}{c|}{ } & 
 \multicolumn{2}{c|}{ } & 
 \multicolumn{2}{c|}{ } \\ 
\hline
\end{tabular}
\\
\vspace*{2cm}  

\begin{minipage}{13cm}
We plot the calculated values of the first six excitation
energies of the boson boson states ($2p-2h$) for four cases of the 
coupling constants with $N=1601$ and $L_0=100$. Here, $n$ of the 
$E_{2p2h}^{(BB)*n}$ denotes the $n-$th energy state from the lowest boson 
boson configurations. 
  
\end{minipage}
\end{center}

\newpage
\begin{center}
\underline{Table 4a} \\
\ \ \\
$\displaystyle{[\frac{g}{\pi}=1]}$ \ \ \ \ \ \\
\ \ \\
\begin{tabular}{|c|c||c|c|c|c|}
\hline
$\rho_0$ & $N$ & $\Delta E^{(BB)*1}_{2p2h}$ & $\Delta E^{(0)}_{1p1h}$ 
& $\Delta E^{(1)}_{1p1h}$ & $\Delta E^{(1,-1)}_{2p2h}$ \\
\hline
4 & 
\begin{tabular}{c} 
101 \\ 201 \\ 401 \\ 1601 
\end{tabular} & 
\begin{tabular}{r@{\extracolsep{0pt}.}l}
 6 & 16 \\
 6 & 06 \\
 6 & 13 \\
 6 & 09 
\end{tabular} & 
\begin{tabular}{r@{\extracolsep{0pt}.}l}
 4 & 47 \\
 4 & 45 \\
 4 & 45 \\
 4 & 44 
\end{tabular} & 
\begin{tabular}{r@{\extracolsep{0pt}.}l}
 9 & 88 \\
 9 & 99 \\
 10 & 05 \\
 10 & 09 
\end{tabular} & 
\begin{tabular}{r@{\extracolsep{0pt}.}l}
 19 & 21 \\
 19 & 71 \\
 19 & 96 \\
 20 & 15 
\end{tabular}  \\
\hline
8 & 
\begin{tabular}{c} 
101 \\ 201 \\ 401 \\ 1601 
\end{tabular} & 
\begin{tabular}{r@{\extracolsep{0pt}.}l}
 5 & 74 \\
 5 & 76 \\
 5 & 79 \\
 5 & 77 
\end{tabular} & 
\begin{tabular}{r@{\extracolsep{0pt}.}l}
 5 & 91 \\
 5 & 89 \\
 5 & 89 \\
 5 & 88 
\end{tabular} & 
\begin{tabular}{r@{\extracolsep{0pt}.}l}
 13 & 93 \\
 14 & 25 \\
 14 & 41 \\
 14 & 53 
\end{tabular} & 
\begin{tabular}{r@{\extracolsep{0pt}.}l}
 26 & 71 \\
 27 & 90 \\
 28 & 52 \\
 28 & 98 
\end{tabular}  \\
\hline
16 & 
\begin{tabular}{c} 

101 \\ 201 \\ 401 \\ 1601 
\end{tabular} & 
\begin{tabular}{r@{\extracolsep{0pt}.}l}
 5 & 55 \\
 5 & 59 \\
 5 & 54 \\
 5 & 57 
\end{tabular} & 
\begin{tabular}{r@{\extracolsep{0pt}.}l}
 7 & 93 \\
 7 & 91 \\
 7 & 89 \\
 7 & 89 
\end{tabular} & 
\begin{tabular}{r@{\extracolsep{0pt}.}l}
 19 & 21 \\
 19 & 94 \\
 20 & 32 \\
 20 & 61 
\end{tabular} & 
\begin{tabular}{r@{\extracolsep{0pt}.}l}
 36 & 03 \\
 38 & 61 \\
 40 & 00 \\
 41 & 07 
\end{tabular}  \\
\hline
32 & 
\begin{tabular}{c} 
101 \\ 201 \\ 401 \\ 1601 
\end{tabular} & 
\begin{tabular}{r@{\extracolsep{0pt}.}l}
 5 & 26 \\
 5 & 30 \\
 5 & 26 \\
 5 & 28 
\end{tabular} & 
\begin{tabular}{r@{\extracolsep{0pt}.}l}
 10 & 72 \\
 10 & 69 \\
 10 & 68 \\
 10 & 67 
\end{tabular} & 
\begin{tabular}{r@{\extracolsep{0pt}.}l}
 25 & 84 \\
 27 & 40 \\
 28 & 23 \\
 28 & 88 
\end{tabular} & 
\begin{tabular}{r@{\extracolsep{0pt}.}l}
 47 & 21 \\
 52 & 14 \\
 55 & 09 \\
 57 & 40 
\end{tabular}  \\
\hline
64 & 
\begin{tabular}{c} 
101 \\ 201 \\ 401 \\ 1601 
\end{tabular} & 
\begin{tabular}{r@{\extracolsep{0pt}.}l}
 4 & 93 \\
 4 & 97 \\
 4 & 94 \\
 $*$ 
\end{tabular} & 
\begin{tabular}{r@{\extracolsep{0pt}.}l}
 14 & 55 \\
 14 & 52 \\
 14 & 51 \\
 14 & 48 
\end{tabular} & 
\begin{tabular}{r@{\extracolsep{0pt}.}l}
 33 & 80 \\
 36 & 95 \\
 38 & 68 \\
 40 & 03 
\end{tabular} & 
\begin{tabular}{r@{\extracolsep{0pt}.}l}
 61 & 06 \\
 68 & 58 \\
 74 & 49 \\
 79 & 28   
\end{tabular}  \\
\hline
\end{tabular}
\\
\vspace*{2cm}  

\begin{minipage}{13cm}
We plot the calculated values of the lowest excitation energy 
of the boson boson states with respect to the vacuum energy 
as the function of $N$ and $\rho_0$ for ${g\over{\pi}}=1$ case. 
Here, we also show the boson energy ($\Delta E_{1p1h}^{(0)}$) 
as well as the lowest excitation energy of the $1p-1h$ 
($\Delta E_{1p1h}^{(1)}$) and $2p-2h$ ($\Delta E_{2p2h}^{(1,-1)}$ 
scattering states with respect to the vacuum energy. Here, $*$ 
shows that the present computor program could not find the solutions. 
\end{minipage}
\end{center}
\newpage

\begin{center}
\underline{Table 4b} \\
\ \ \\
$\displaystyle{[\frac{g}{\pi}=1.25]}$ \ \ \ \ \ \\
\ \ \\
\begin{tabular}{|c|c||c|c|c|c|}
\hline
$\rho_0$ & $N$ & $\Delta E^{(BB)*1}_{2p2h}$ & $\Delta E^{(0)}_{1p1h}$ 
& $\Delta E^{(1)}_{1p1h}$ & $\Delta E^{(1,-1)}_{2p2h}$ \\
\hline
4 & 
\begin{tabular}{c} 
101 \\ 201 \\ 401 \\ 1601 
\end{tabular} & 
\begin{tabular}{r@{\extracolsep{0pt}.}l}
 5 & 70 \\
 5 & 61 \\
 5 & 61 \\
 5 & 61 
\end{tabular} & 
\begin{tabular}{r@{\extracolsep{0pt}.}l}
 3 & 84 \\
 3 & 83 \\
 3 & 82 \\
 3 & 82 
\end{tabular} & 
\begin{tabular}{r@{\extracolsep{0pt}.}l}
11 & 79 \\
12 & 09 \\
 12 & 25 \\
 12 & 37 
\end{tabular} & 
\begin{tabular}{r@{\extracolsep{0pt}.}l}
 23 & 19 \\
 24 & 00 \\
 24 & 41 \\
 24 & 71 
\end{tabular}  \\
\hline
8 & 
\begin{tabular}{c} 
101 \\ 201 \\ 401 \\ 1601 
\end{tabular} & 
\begin{tabular}{r@{\extracolsep{0pt}.}l}
 5 & 68 \\
 5 & 71 \\
 5 & 72 \\
 5 & 75 
\end{tabular} & 
\begin{tabular}{r@{\extracolsep{0pt}.}l}
 5 & 02 \\
 5 & 02 \\
 5 & 01 \\
 5 & 01 
\end{tabular} & 
\begin{tabular}{r@{\extracolsep{0pt}.}l}
 17 & 12 \\
 17 & 90 \\
 18 & 30 \\
 18 & 62 
\end{tabular} & 
\begin{tabular}{r@{\extracolsep{0pt}.}l}
 33 & 32 \\
 35 & 34 \\
 36 & 39 \\
 37 & 18 
\end{tabular}  \\
\hline
16 & 
\begin{tabular}{c} 
101 \\ 201 \\ 401 \\ 1601 
\end{tabular} & 
\begin{tabular}{r@{\extracolsep{0pt}.}l}
 5 & 68 \\
 5 & 72 \\
 5 & 74 \\
 5 & 73 
\end{tabular} & 
\begin{tabular}{r@{\extracolsep{0pt}.}l}
 6 & 75 \\
 6 & 74 \\
 6 & 73 \\
 6 & 72 
\end{tabular} & 
\begin{tabular}{r@{\extracolsep{0pt}.}l}
 24 & 02 \\
 25 & 76 \\
 26 & 69 \\
 27 & 42 
\end{tabular} & 
\begin{tabular}{r@{\extracolsep{0pt}.}l}
 45 & 68 \\
 50 & 39 \\
 52 & 86 \\
 54 & 73 
\end{tabular}  \\
\hline
32 & 
\begin{tabular}{c} 
101 \\ 201 \\ 401 \\ 1601 
\end{tabular} & 
\begin{tabular}{r@{\extracolsep{0pt}.}l}
 5 & 57 \\
 5 & 62 \\
 5 & 58 \\
 $*$ 
\end{tabular} & 
\begin{tabular}{r@{\extracolsep{0pt}.}l}
  9 & 21 \\
  9 & 19 \\
  9 & 18 \\
  9 & 18 
\end{tabular} & 
\begin{tabular}{r@{\extracolsep{0pt}.}l}
 32 & 44 \\
 36 & 04 \\
 38 & 05 \\
 39 & 65 
\end{tabular} & 
\begin{tabular}{r@{\extracolsep{0pt}.}l}
 59 & 79 \\
 69 & 38 \\
 74 & 82 \\
 79 & 00 
\end{tabular}  \\
\hline
64 & 
\begin{tabular}{c} 
101 \\ 201 \\ 401 \\ 1601 
\end{tabular} & 
\begin{tabular}{r@{\extracolsep{0pt}.}l}
 5 & 40 \\
 $*$ \\
 $*$ \\
 $*$ 
\end{tabular} & 
\begin{tabular}{r@{\extracolsep{0pt}.}l}
 12 & 65 \\
 12 & 62 \\
 12 & 62 \\
 12 & 60 
\end{tabular} & 
\begin{tabular}{r@{\extracolsep{0pt}.}l}
 42 & 00 \\
 48 & 98 \\
 53 & 07 \\
  $*$ 
\end{tabular} & 
\begin{tabular}{r@{\extracolsep{0pt}.}l}
 75 & 57 \\
 91 & 89 \\
103 & 18 \\
  $*$    
\end{tabular}  \\
\hline
\end{tabular}
\\
\vspace*{2cm}  
\begin{minipage}{13cm}
The same as Table 4a. The coupling constant is ${g\over{\pi}}=1.25$.
\end{minipage}
\end{center}
\newpage
\begin{center}
\underline{Table 4c} \\
\ \ \\
$\displaystyle{[\frac{g}{\pi}=1.5]}$ \ \ \ \ \ \\
\ \ \\
\begin{tabular}{|c|c||c|c|c|c|}
\hline
$\rho_0$ & $N$ & $\Delta E^{(BB)*1}_{2p2h}$ & $\Delta E^{(0)}_{1p1h}$ 
& $\Delta E^{(1)}_{1p1h}$ & $\Delta E^{(1,-1)}_{2p2h}$ \\
\hline
4 & 
\begin{tabular}{c} 
101 \\ 201 \\ 401 \\ 1601 
\end{tabular} & 
\begin{tabular}{r@{\extracolsep{0pt}.}l}
 5 & 43 \\
 5 & 31 \\
 5 & 37 \\
 5 & 33 
\end{tabular} & 
\begin{tabular}{r@{\extracolsep{0pt}.}l}
 3 & 21 \\
 3 & 21 \\
 3 & 21 \\
 3 & 21 
\end{tabular} & 
\begin{tabular}{r@{\extracolsep{0pt}.}l}
13 & 72 \\
14 & 29 \\
 14 & 59 \\
 14 & 81 
\end{tabular} & 
\begin{tabular}{r@{\extracolsep{0pt}.}l}
 27 & 33 \\
 28 & 57 \\
 29 & 18 \\
 29 & 61 
\end{tabular}  \\
\hline
8 & 
\begin{tabular}{c} 
101 \\ 201 \\ 401 \\ 1601 
\end{tabular} & 
\begin{tabular}{r@{\extracolsep{0pt}.}l}
 5 & 53 \\
 5 & 55 \\
 5 & 62 \\
 5 & 59 
\end{tabular} & 
\begin{tabular}{r@{\extracolsep{0pt}.}l}
 4 & 08 \\
 4 & 08 \\
 4 & 07 \\
 4 & 07 
\end{tabular} & 
\begin{tabular}{r@{\extracolsep{0pt}.}l}
 20 & 40 \\
 21 & 88 \\
 22 & 67 \\
 23 & 29 
\end{tabular} & 
\begin{tabular}{r@{\extracolsep{0pt}.}l}
 40 & 30 \\
 43 & 67 \\
 45 & 34 \\
 46 & 58 
\end{tabular}  \\
\hline
16 & 
\begin{tabular}{c} 
101 \\ 201 \\ 401 \\ 1601 
\end{tabular} & 
\begin{tabular}{r@{\extracolsep{0pt}.}l}
 5 & 71 \\
 5 & 74 \\
 5 & 84 \\
 $*$ 
\end{tabular} & 
\begin{tabular}{r@{\extracolsep{0pt}.}l}
 5 & 41 \\
 5 & 40 \\
 5 & 39 \\
 5 & 39 
\end{tabular} & 
\begin{tabular}{r@{\extracolsep{0pt}.}l}
 28 & 96 \\
 32 & 29 \\
 34 & 16 \\
 35 & 65 
\end{tabular} & 
\begin{tabular}{r@{\extracolsep{0pt}.}l}
 55 & 88 \\
 63 & 97 \\
 68 & 15 \\
 71 & 28 
\end{tabular}  \\
\hline
32 & 
\begin{tabular}{c} 
101 \\ 201 \\ 401 \\ 1601 
\end{tabular} & 
\begin{tabular}{r@{\extracolsep{0pt}.}l}
 5 & 74 \\
 $*$ \\
 $*$ \\
 $*$
\end{tabular} & 
\begin{tabular}{r@{\extracolsep{0pt}.}l}
  7 & 34 \\
  7 & 33 \\
  $*$ \\
  $*$ 
\end{tabular} & 
\begin{tabular}{r@{\extracolsep{0pt}.}l}
 39 & 02 \\
 45 & 82 \\
  $*$ \\
  $*$ 
\end{tabular} & 
\begin{tabular}{r@{\extracolsep{0pt}.}l}
 72 & 31 \\
 89 & 22 \\
  $*$ \\
  $*$ 
\end{tabular}  \\
\hline
64 & 
\begin{tabular}{c} 
101 \\ 201 \\ 401 \\ 1601 
\end{tabular} & 
\begin{tabular}{r@{\extracolsep{0pt}.}l}
 $*$ \\
 $*$ \\
 $*$ \\
 $*$ 
\end{tabular} & 
\begin{tabular}{r@{\extracolsep{0pt}.}l}
  $*$ \\
  $*$ \\
  $*$ \\
  $*$ 
\end{tabular} & 
\begin{tabular}{r@{\extracolsep{0pt}.}l}
  $*$ \\
  $*$ \\
  $*$ \\
  $*$ 
\end{tabular} & 
\begin{tabular}{r@{\extracolsep{0pt}.}l}
  $*$ \\
  $*$ \\
  $*$ \\
  $*$    
\end{tabular}  \\
\hline
\end{tabular}
\\
\vspace*{2cm}  
\begin{minipage}{13cm}
The same as Table 4a. The coupling constant is ${g\over{\pi}}=1.5$.
\end{minipage}
\end{center}
\newpage
\begin{center}
\underline{Table 4d} \\
\ \ \\
$\displaystyle{[\frac{g}{\pi}=1.7]}$ \ \ \ \ \ \\
\ \ \\
\begin{tabular}{|c|c||c|c|c|c|}
\hline
$\rho_0$ & $N$ & $\Delta E^{(BB)*1}_{2p2h}$ & $\Delta E^{(0)}_{1p1h}$ 
& $\Delta E^{(1)}_{1p1h}$ & $\Delta E^{(1,-1)}_{2p2h}$ \\
\hline
4 & 
\begin{tabular}{c} 
101 \\ 201 \\ 401 \\ 1601 
\end{tabular} & 
\begin{tabular}{r@{\extracolsep{0pt}.}l}
 5 & 10 \\
 5 & 07 \\
 4 & 99 \\
 4 & 98 
\end{tabular} & 
\begin{tabular}{r@{\extracolsep{0pt}.}l}
 2 & 73 \\
 2 & 73 \\
 2 & 73 \\
 2 & 72 
\end{tabular} & 
\begin{tabular}{r@{\extracolsep{0pt}.}l}
15 & 46 \\
16 & 35 \\
 16 & 82 \\
 17 & 17 
\end{tabular} & 
\begin{tabular}{r@{\extracolsep{0pt}.}l}
 31 & 19 \\
 32 & 93 \\
 33 & 78 \\
 34 & 38 
\end{tabular}  \\
\hline
8 & 
\begin{tabular}{c} 
101 \\ 201 \\ 401 \\ 1601 
\end{tabular} & 
\begin{tabular}{r@{\extracolsep{0pt}.}l}
 5 & 24 \\
 5 & 41 \\
 5 & 30 \\
 5 & 31 
\end{tabular} & 
\begin{tabular}{r@{\extracolsep{0pt}.}l}
 3 & 31 \\
 3 & 31 \\
 3 & 31 \\
 3 & 30 
\end{tabular} & 
\begin{tabular}{r@{\extracolsep{0pt}.}l}
 23 & 37 \\
 25 & 81 \\
 27 & 14 \\
 28 & 18 
\end{tabular} & 
\begin{tabular}{r@{\extracolsep{0pt}.}l}
 46 & 94 \\
 52 & 11 \\
 54 & 65 \\
 56 & 49 
\end{tabular}  \\
\hline
16 & 
\begin{tabular}{c} 
101 \\ 201 \\ 401 \\ 1601 
\end{tabular} & 
\begin{tabular}{r@{\extracolsep{0pt}.}l}
 5 & 53 \\
 6 & 01 \\
 $*$ \\
 $*$ 
\end{tabular} & 
\begin{tabular}{r@{\extracolsep{0pt}.}l}
 4 & 24 \\
 4 & 22 \\
 $*$ \\
 $*$ 
\end{tabular} & 
\begin{tabular}{r@{\extracolsep{0pt}.}l}
 33 & 29 \\
 38 & 87 \\
  $*$ \\
  $*$  
\end{tabular} & 
\begin{tabular}{r@{\extracolsep{0pt}.}l}
 64 & 98 \\
 78 & 09 \\
  $*$  \\
  $*$  
\end{tabular}  \\
\hline
32 & 
\begin{tabular}{c} 
101 \\ 201 \\ 401 \\ 1601 
\end{tabular} & 
\begin{tabular}{r@{\extracolsep{0pt}.}l}
 $*$  \\
 $*$  \\
 $*$ \\
 $*$
\end{tabular} & 
\begin{tabular}{r@{\extracolsep{0pt}.}l}
  5 & 64 \\
  $*$  \\
  $*$ \\
  $*$ 
\end{tabular} & 
\begin{tabular}{r@{\extracolsep{0pt}.}l}
 44 & 38 \\
  $*$  \\
  $*$ \\
  $*$ 
\end{tabular} & 
\begin{tabular}{r@{\extracolsep{0pt}.}l}
 82 & 24 \\
  $*$  \\
  $*$ \\
  $*$ 
\end{tabular}  \\
\hline
64 & 
\begin{tabular}{c} 
101 \\ 201 \\ 401 \\ 1601 
\end{tabular} & 
\begin{tabular}{r@{\extracolsep{0pt}.}l}
 $*$ \\
 $*$ \\
 $*$ \\
 $*$ 
\end{tabular} & 
\begin{tabular}{r@{\extracolsep{0pt}.}l}
  $*$ \\
  $*$ \\
  $*$ \\
  $*$ 
\end{tabular} & 
\begin{tabular}{r@{\extracolsep{0pt}.}l}
  $*$ \\
  $*$ \\
  $*$ \\
  $*$ 
\end{tabular} & 
\begin{tabular}{r@{\extracolsep{0pt}.}l}
  $*$ \\
  $*$ \\
  $*$ \\
  $*$    
\end{tabular}  \\
\hline
\end{tabular}
\\
\vspace*{2cm}  
\begin{minipage}{13cm}
The same as Table 4a. The coupling constant is ${g\over{\pi}}=1.7$.
\end{minipage}
\end{center}

\newpage
Figure captions:

\begin{list}{}{}
\item[Fig.1:] In fig.1a, we show the configuration of the vacuum in the 
$E-k$ plane. Fig.1b shows the symmetric case of $1p-1h$ state 
while fig.1c the asymmetric case of $1p-1h$ state. 
Fig.1d shows the symmetric case of $2p-2h$ state. 

\item[Fig.2:] The calculated spectrum for ${g\over{\pi}}=1.25$ with $L_0=110$ 
and $N=1601$ is shown. The ``boson" corresponds to $\Delta E_{1p1h}^{(0)}$ 
while all the other states are in the continuum. We identify the physical 
fermion mass such that the lowest energy of the $1p-1h$ continuum state is 
$2m$. 

\item[Fig.3:] We show  the excitation energies $\Delta E_{1p1h}^{(0)}$  
and  $\Delta E_{1p1h}^{(1)}$   for 
 the coupling constant ${g\over{\pi}}=0.8$ ($g_0 =4.19$) (fig.3a),   
  ${g\over{\pi}}=1.0$ ($g_0 =6.28$) (fig.3b)  and 
    ${g\over{\pi}}=1.25$ ($g_0 =10.5$) (fig.3c) cases 
 as the function of $\rho_0$. Here, $E(0)$ and $E(1)$ denote 
 $\Delta E_{1p1h}^{(0)}$  and  $\Delta E_{1p1h}^{(1)}$, respectively. 
The black dots denote the calculated results. The solid lines are 
straight lines for reference 
in log-log plot with the same slope $\alpha$ for 
 $\Delta E_{1p1h}^{(0)}$  and  $\Delta E_{1p1h}^{(1)}$.      
 
\item[Fig.4:] The same excitation energies  $\Delta E_{1p1h}^{(0)}$  
and  $\Delta E_{1p1h}^{(1)}$  are plotted as the function of 
$\rho^{\alpha}_0$. The numbers on the lines denote 
the coupling constant ${g\over{\pi}}$.  The black circles are for 
$\Delta E_{1p1h}^{(0)}$  while the black squares 
 for  $\Delta E_{1p1h}^{(1)}$. The dashed lines 
are straight line for reference. 

\item[Fig.5:] The boson mass is shown as the function of the 
coupling constant ${g\over{\pi}}$. 
The black circles with error bars ( $\Delta {\cal M}$ ) are the present 
calculation. The solid line ( FO ) is the predicted boson mass by 
Fujita-Ogura, the dashed line ( DHN) by Dashen et al. and 
the dashed-dotted line ( BT ) by Bergknoff and Thacker.   

\item[Fig.6:] We show the calculated excitation energies 
$\Delta E_{1p1h}^{(0)}$  and  $\Delta E_{1p1h}^{(1)}$  
for higher coupling constants ( ${g\over{\pi}}=$ 1.5 and 1.7 ) as 
the function of $\rho_0$. The black dots denote for ${g\over{\pi}}=$ 1.5 
 while the white circles for ${g\over{\pi}}=$ 1.7. 
As can be seen, the slopes of the 
two excitation energies are not the same with each other.

\end{list}
\end{document}